# Can rotation curves reveal the opacity of spiral galaxies ?

Angelos Misiriotis

*University of Crete, Physics Department, P.O. Box 2208, 710 03 Heraklion, Crete, Greece*

**Abstract.** Long-slit optical spectra along the major axis of spiral galaxies constitute a main source of information concerning the kinematics of galaxies. In highly inclined spiral galaxies the long-slit spectrum is distorted by dust, which obscures the innermost parts of the galaxy and eliminates the relevant kinematic information. This effect however may provide information concerning the amount and the spatial distribution of dust. Here I review previous attempts to quantify the internal extinction of spiral galaxies using kinematic information and I discuss the relevant theoretical models. Finally I suggest possible directions for further studies on this subject.

## INTRODUCTION

Long-slit optical spectra along the major axis of spiral galaxies constitute a main source of information concerning the kinematics of galaxies (see [1] for a review). For highly inclined spiral galaxies however the long-slit spectrum is distorted by dust, which obscures the innermost parts of the galaxy and eliminates the relevant kinematic information. As first noted by Goad & Roberts ([2]), a sufficiently optically thick spiral disk would yield a solid-body rotation curve regardless of the shape of the actual rotation curve. Therefore, the shape of the rotation curves of edge-on galaxies may be used as a diagnostic for their opacity. Dust however is not the only cause for the distortion of the observed rotation curve. To be able to draw quantitative conclusions for the opacity, all the relevant factors should be taken into account.

In the hypothetical complete absence of dust, when the inclination is high, any line of sight crosses a large part of the galaxy. As a result, the total observed spectral line consists of kinematical information from many radii in the galaxy. This is illustrated in Fig. 1. The right panel shows the top view of a galaxy. Here the rotational velocity of each emitter in the galaxy depends only on the distance from the center. The galaxy is divided in different zones indicated in the plot with different filling-patterns. The arrows, indicate the velocity at each zone. If we assume an observer which is located on the galactic plane and observes the galaxy from $y = -\infty$, then all the lines of sight for this observer are parallel to the $y$ axis. Of course, this observer sees the galaxy edge-on. The shaded rectangle shows one of the lines of sight which goes through the whole galaxy "collecting" kinematical information from many different radii. In the left panel of Fig. 1 the total observed spectral line profile is shown (thick solid line). This profile comes from the contribution of the various zones of the galaxy. Each contribution is shaded according to the zone that is responsible for its emission. Note that, even though the outer parts of the galaxy exhibit the same rotational velocity, their contribution to the line-of sight appears at smaller and smaller velocities as the radius increases due to the projection of the rotational velocity on the line of sight. Note also that the profile is asymmetric, and therefore its intensity weighted average value does not correspond to the position of the maximum.

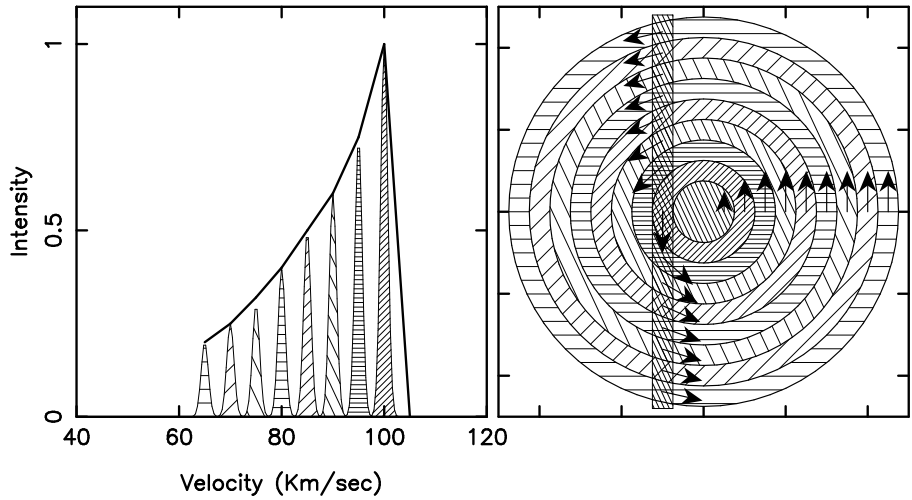

**FIGURE 1.** Illustration of the observed spectral line (left panel) originating from an edge-on line of sight (right panel) and collects kinematical information from the parts indicated by the shaded rectangle

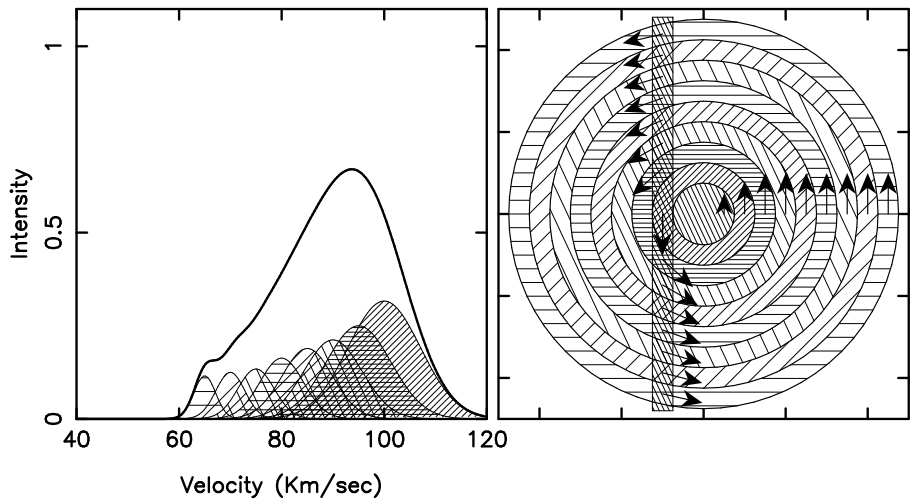

**FIGURE 2.** Same as in Fig. 1 but with the inclusion of random motion for the emitters.

Let us now take into account that the rotation of the emitters around the galaxy is not perfectly circular but in addition to the mean rotational velocity there are random motions characterized by a velocity dispersion. When the velocity dispersion is not negligible, the contribution of each zone to the total observed line profile is not a sharp peak but an extended Gaussian distribution. The effect of the velocity dispersion can be seen in Fig. 2. The right panel is the same as in Fig. 1, but in the left panel one can see that the individual contribution of each zone is no longer a sharp peak. The total observed line profile (indicated with the thick line) is obviously unsharpened and its maximum is pushed towards lower velocities [1].

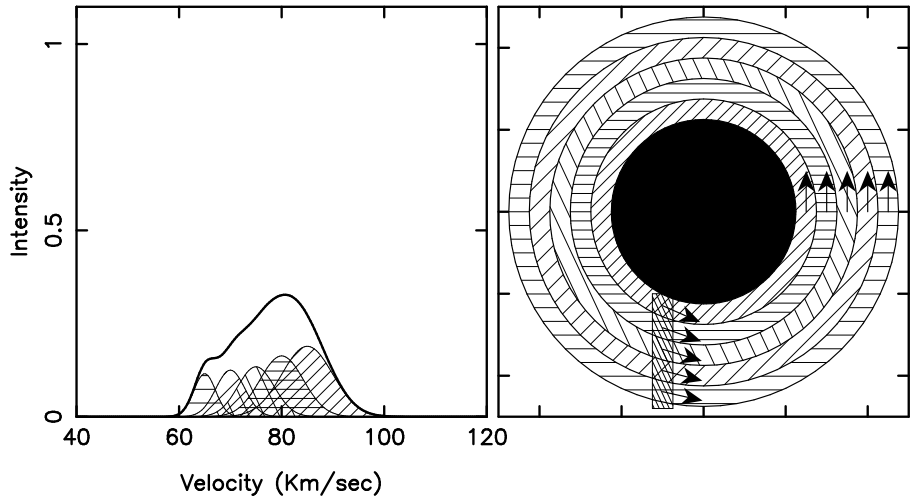

**FIGURE 3.** Same as in Fig. 2 but with the inner parts of the galaxy obscured by dust.

In Fig. 3 the effect of dust is shown. In the right panel, the black filled circle indicates the part of the galaxy which due to dust is impenetrable to the edge-on observer. The kinematical information observed originates from the outer parts of the galaxy while the contribution of the innermost parts is eliminated. The result can be seen in the left panel. Since the high-velocity contribution from the innermost parts is gone, the total profile is pushed towards lower velocities. A second very important effect is that the broadening of the total observed profile is now smaller.

## THEORETICAL MODELS

To quantify the effects mentioned in the previous section one has to rely on a model. The minimum set of parameters for such a model should describe

- the kinematics of the emitters,
- the spatial distribution of the emitters,
- the spatial distribution of the dust.

In addition to these, if one wants to take properly into account scattering by dust, then the kinematics of the dust should be also included. There is a small number of such models in the literature. The most recent and complete work can be found in Baes et al. ([5]) who studied numerically the effect of dust on the stellar kinematics using a Monte Carlo radiative transfer code and properly included random motions of the stars and the dust as well as scattering. The reader is also referred to the works of Matthews & Wood ([6]), Zasov & Khoperskov ([7]), Valotto & Giovanelli ([8]), and Kregel & van der Kruit ([9]) for related calculations of varying sophistication.

---

[1] It should be emphasized that in this case, the intrinsic rotational velocity can be derived accurately through the use of methods more sophisticated that the intensity weighted average (e.g. [3]; [4]).

## OBSERVATIONS

An obvious method to probe the effect of dust on the kinematics is the comparison of kinematics in different wavelengths. This approach was first adopted by Bosma et al. ([10]), who compared the optical rotation curve of NGC 891 with its 21cm and CO line kinematics. They concluded that NGC 891 might be moderately opaque at its central regions, but is definitely transparent at the outer regions. Following this work, Bosma ([11]) presented further similar observations on a number of edge-on galaxies. His results (under the assumption of an exponential distribution for the dust) imply that the central face-on optical depth of spiral galaxies at H$\alpha$ is of the order of unity at most.

Along the same lines, Prada et al. ([12]) conducted long-slit spectroscopy on NGC 2146 at two different wavelengths (H$\alpha$ and [NIII]) to conclude that while the extinction is significant on the dust lane, it is limited at the outer parts of the disk.

Kregel et al. ([13]) studied the stellar kinematics of 17 edge-on galaxies and found that away from the dust lane the rotation curves are far from solid body; thus the effect of extinction is not large. In the cases however where the long slit spectrum was taken perpendicular to the major axis, one can see the signature of the extinction as a drop of the velocity and the velocity dispersion profile. Kregel & van der Kruit ([9]) found that in the case of NGC 891 the drop is consistent with a value of the face-on optical depth in the V-band lower than unity, supporting the results of Xilouris et al. ([14]) who studied NGC 891 and concluded that it's face-on optical depth is less than one in the V-band.

Giovanelli & Haynes ([15]) applied the Goad & Roberts ([2]) test in a statistical fashion on a sample of some thousand spiral galaxies and found that there is an anti-correlation between the inclination of spiral galaxies and the slope of the inner part of their H$\alpha$ rotation curves. This correlation is most prominent in luminosity-class I and II spirals. Valloto & Giovanelli ([8]) interpreted this correlation as an effect solely due to dust absorption and argued that the luminosity-class I subsample exhibits an average central face-on optical depth $\tau$ at H$\alpha$ equal to 2 while for the luminosity class II subsample their quoted value was equal to 3.5. Since several recent optical studies ([14],[16] and references therein) confirm that the central optical depth of luminous spiral galaxies is less than unity in all optical bands this appears as a controversy. I will show that this controversy can be resolved with the adoption of a more elaborate model for the distribution of the emitters than the standard exponential model.

| $h_{H\alpha}$ | $z_{H\alpha}$ | $\beta$ | $h_d$ | $z_d$ | $V_{max}$ | $h_r$ |
| kpc | kpc | | kpc | kpc | km/sec | kpc |
|---|---|---|---|---|---|---|
| 8.0 | 0.4 | 1 | 8.0 | 0.4 | 100 | 1.8 |

**TABLE 1.** Parameters describing the model

# MODEL

The model that I use consists of an axisymmetric exponential disk with a central suppression ([17]) for the H$\alpha$ emitters whose emissivity is given by

$$\eta = \eta_0 \left(\frac{R}{h_{H\alpha}}\right)^\beta \exp(-R/h_{H\alpha})\exp(-|z|/z_{H\alpha}), \tag{1}$$

where $\eta_0$ is the emissivity normalization constant and is set equal to 1 throughout the rest of the paper. Here $R$ and $z$ are the cylindrical coordinates, $h_{H\alpha}$ is the disk scalelength and $z_{H\alpha}$ the disk scaleheight. Finally, the central suppression of the distribution is described by $\beta$.

The model also consists of an exponential disk for the dust whose extinction coefficient is given by

$$\kappa = \kappa_0 \exp(-R/h_d - |z|/z_d), \tag{2}$$

where $\kappa_0$ is the extinction coefficient at H$\alpha$ at the center of the disk, $h_d$ is the disk scalelength and $z_d$ is the disk scaleheight. The constant $\kappa_0$ is linked to the central face-on optical depth through $\tau = 2\kappa_0 z_d$.

The mean rotational speed of the emitters in the disk is given by a flat "Polyex" model (GH02), where the rotational velocity $V(R)$ as a function of radius is given by

$$V(R) = V_{max}[1 - \exp(-R/h_r)], \tag{3}$$

where $V_{max}$ is the velocity at the flat part of the rotation curve (taken 100 km s$^{-1}$ throughout the rest of the paper), and $h_r$ is the scalelength of the rotation curve.

In addition to the mean rotational velocity, the emitters exhibit random motions described by the radial, the azimuthal, and the vertical velocity dispersions, $\sigma_R$, $\sigma_\phi$, and $\sigma_z$, respectively. Throughout the rest of the paper we will assume that the velocity dispersion in all directions is 10 km s$^{-1}$ in accordance with the fact that the gas is dynamically cold.

Unfortunately, not only the number of parameters in the above model is quite large, but also the resulting long-slit spectrum (and as a consequence the shape of the rotation curve) depends strongly on each one of them. However, the adoption of a set of reasonable values can yield a model which describes realistically, if not accurately, an "average" spiral galaxy. The set adopted in this work is presented in Table 1 and it is discussed in further detail in [18].

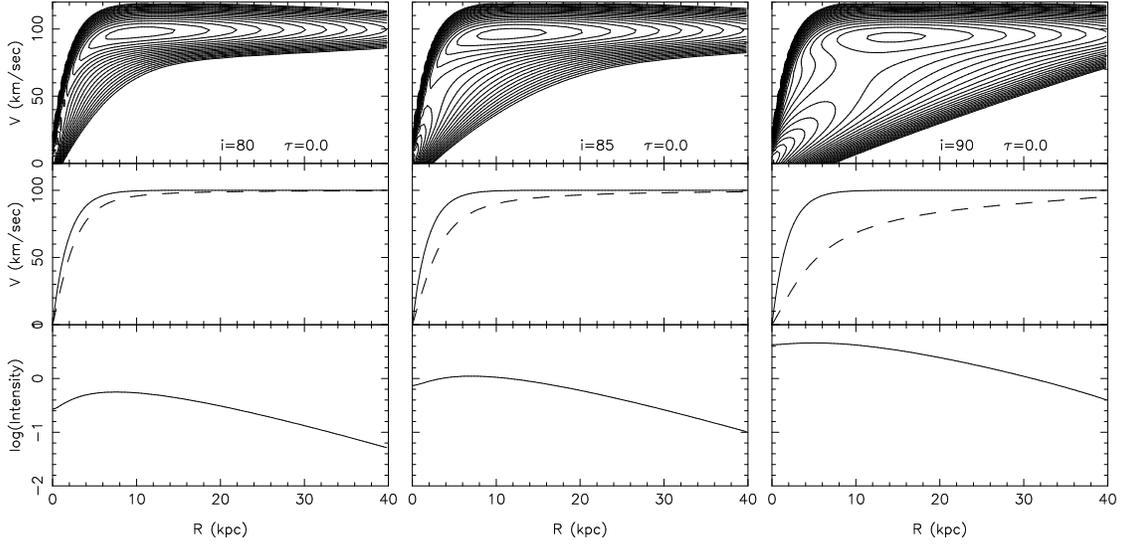

**FIGURE 4.** Long slit spectra of a model without dust. The first column corresponds to inclination 80°, the second column to 85° and the third column to 90°. The top panel show the long slit spectra. Isocontours are drawn logarithmically so that ten of them correspond to one order of magnitude. The middle panel shows the intrinsic (solid line) and the derived (dashed line) rotation curve. The bottom panel shows the intensity along the semi-major axis.

## Optically thin case

In the first column of Fig. 4 at the top panel I show the simulated long-slit spectrum of a model with inclination angle 80° and no dust. The spectrum follows very well the the intrinsic rotation curve and allows its accurate reproduction even using the simplest technique of the weighted mean of the spectral line which I use here. I demonstrate this fact in the middle panel of the first column, where I plot the intrinsic rotation curve (solid line) and the derived rotation curve (dashed line). The small disagreement originates from the fact that the spectral line profile of the simulated long slit spectrum is not symmetrical but exhibits a longer tail towards lower velocities, thus resulting in a somewhat lower value for the mean (see [19] for an extended discussion on this subject). This is widely referred to as "projection effect". In the bottom panel I show in semilogarithmic scale the major axis profile of the intensity. The central trough originates from the suppression in the gas distribution I adopted.

In the second column at the top panel I plot the long-slit spectrum for an inclination of 85°. The higher inclination increases the projection effect and as a consequence, the derived velocity curve in the middle panel of the second column differs from the intrinsic. It should be noted however, that a more sophisticated method would make it possible to retrieve the original rotation curve. In the bottom panel, the central trough at the major axis profile of the intensity is less prominent since the higher inclination causes an increased contribution of the emitters along the line of sight.

In the third column the edge-on case with inclination 90° is shown. Here the projection effects are severe causing a significant difference between the intrinsic and the derived rotation curve. It must be noted here that the relative absence of atomic gas in the central parts of the galaxy mimics absorption by dust there. The inner parts of the galaxy are not represented in the long slit spectrum and consequently in the rotation curve. Still, the intrinsic rotation curve can be derived accurately from the long slit spectrum through the use of a more sophisticated method (e.g. [3], [4]). In the bottom panel, the central drop is almost gone, giving its place to a plateau.

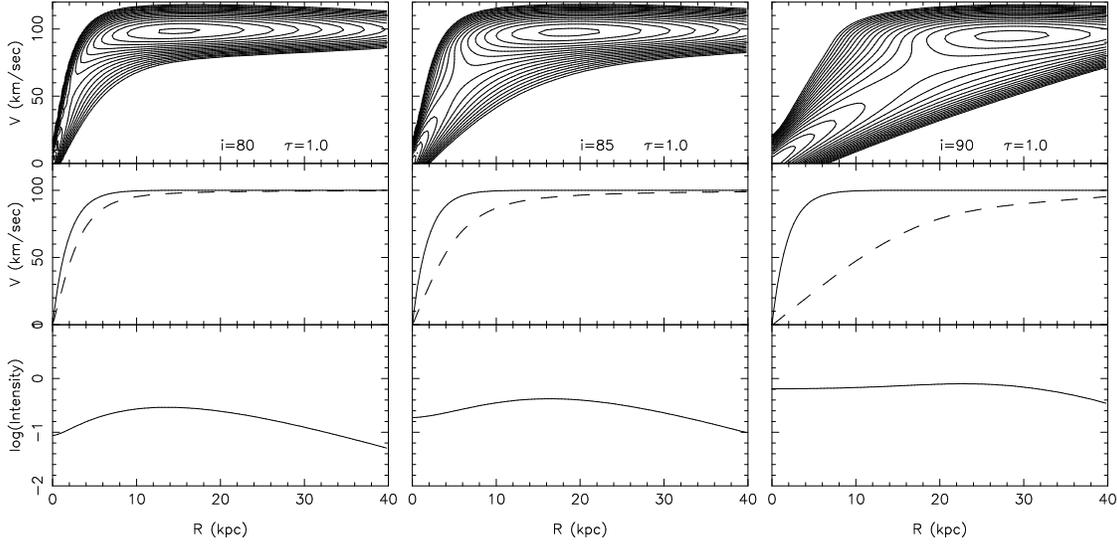

**FIGURE 5.** The same as in Fig. 3 but for central face-on optical depth $\tau = 1$ at H$\alpha$.

## Moderately optically thick case ($\tau = 1$)

In the first column of Fig. 5 at the top panel I show the simulated long-slit spectrum of a model with inclination angle 80° and central face-on optical depth at H$\alpha$ equal to 1. A comparison of this long slit spectrum with the long slit spectrum of the dustless model shows that the effects of the dust are rather limited. In the middle panel I plot the intrinsic (solid line) and the derived (dashed) rotation curve. It is obvious that the small difference is attributed to projection effects rather that to dust extinction. Even in the presence of dust, a sophisticated method would be able to retrieve the intrinsic rotation curve. The effect of the dust is more obvious in the bottom panel where the intensity at the central part of the major axis profile is reduced to about 30% of its intrinsic value. In the second column, where the inclination is 85°, again most of the difference between the intrinsic and the derived rotation curve originates from projection effects. This however does not mean that there is no extinction. In the bottom panel, where the intensity along the major axis is plotted, the absorption is very prominent as the derived intensity at R=0 is reduced to 25% of its original value. Finally in the third column where I show the edge on case (90°) the effects of the dust become prominent. There the difference of the derived rotation curve from the intrinsic one is more due to dust than to projection effect. Even the use of sophisticated techniques in retrieving the intrinsic rotation curve from the long slit spectrum would fail, as the kinematic information from the innermost

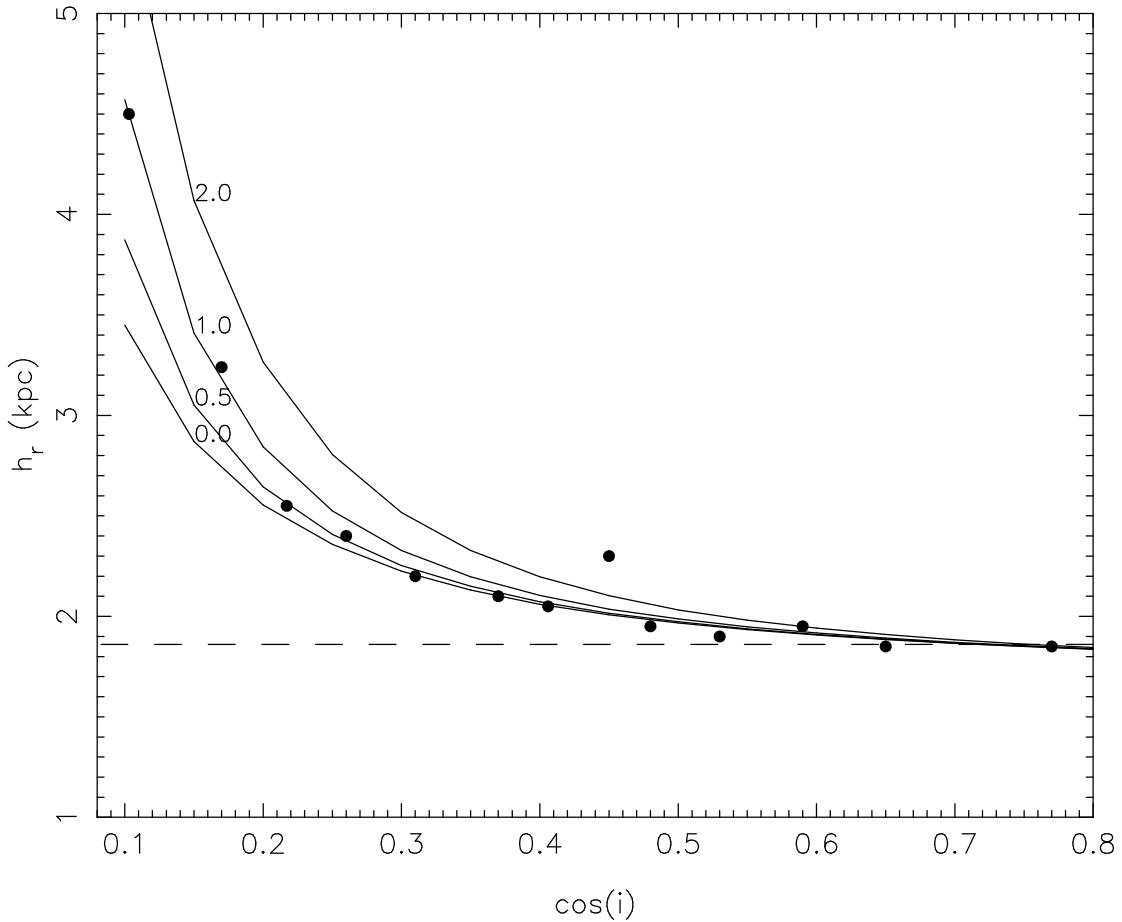

**FIGURE 6.** The observed rotation curve characteristic scale $h_r$ as a function of inclination. The solid circles are the data from [8] while the solid lines show the model results for several different central face-on optical depths. The data points fall roughly between the $\tau=0$ and the $\tau=1$ models. The horizontal line at 1.8 shows the intrinsic value of $h_r$.

parts of the galaxy is absent.

## THE INCLINATION-ROTATION CURVE SHAPE CORRELATION AND THE OPACITY OF SPIRAL GALAXIES.

Since the existence of dust and/or velocity dispersion causes the rotation curves of highly inclined galaxies to rise less steeply than in dustless galaxies of the same inclination, the relation between the inclination angle and the slope of the inner rotation curve can be used as a diagnostic for their opacity. A statistical sample of spiral galaxies was presented by GH02 that shows such a correlation. In [8], Valotto & Giovanelli argued that the implied average central face-on optical depth of luminosity-class I galaxies at H$\alpha$ is around 2.5 while for luminosity-class II it is around 3.5.

Using the model presented above, I calculate the observed $h_r$ as a function of inclination. In Fig. 6 I show $h_r$ as a function of inclination for four different models exhibiting $\tau$=0, 0.5, 1, and 2. The solid circles show the data from [8]. Even for $\tau$=0 there is a significant increase of $h_r$ as the galaxy approaches the edge-on view. As the amount of dust increases, the effects on the rotation curve become more severe and push $h_r$ to higher values. The number of parameters of the model does not allow me to infer the exact central face-on optical depth. Yet a comparison of the model curves with the data allows one to safely state that the central face-on optical depth lies between 0 and 1. This is at least a factor of two lower than the value of 2.5 derived by Valotto & Giovanelli ([8]) for luminosity-class I galaxies and in accordance with recent studies of the opacity. Note that even for dustless galaxies the projection effects alone are enough to create a significant part of the observed correlation.

## DISCUSSION AND CONCLUSIONS

As it has been already noted, it is nearly impossible to quote an exact value for the central face-on optical depth at H$\alpha$. This is due to the fact that the results are sensitive to the following factors:

1. The adopted scaleheight to scalelength ratio $q_{H\alpha}$ of the H$\alpha$ emitters and the dust, which was assumed equal to 0.05. As it is also noted in [8], the results are very sensitive to this ratio.
2. The adopted velocity dispersion $\sigma_0$, for which I have adopted the value of 10 km/sec. Had the velocity dispersion been higher, the dustless observed rotation curve would be flatter (see [7] for an extensive discussion). The additional flattening due to dust would be less and as a result the derived amount of dust would be less.
3. The assumed H$\alpha$ distribution especially in the central parts of the galaxy plays a vital role on the magnitude of the effects of the dust. In addition to the fact that a very large percentage of spiral galaxies exhibit at least a weak bar the validity of the assumed H$\alpha$ axisymmetric distribution should be tested. Such a test however requires observations and detailed modeling of individual galaxies.

The conclusions of this study can be summarized as follows. Using an innerly truncated exponential model for the spatial distribution of the H$\alpha$ emission and a simple exponential for the distribution of the dust in spiral galaxies I reproduce the data of [15] that show a correlation between the inclination angle and the slope of the inner rotation curve of luminosity class I spiral galaxies. My analysis shows that the average central face-on optical depth of luminosity-class I galaxies from the sample of [15] is between 0 and 1. This result is in line with the recent study of Xilouris et al. ([16]).

The problem of the opacity of spiral galaxies seemed to find its solution after extended observations in the SCUBA bands (e.g. [20], [21]) where the bulk of the dust of a galaxy can be imaged. However, detailed modeling to confirm the energy balance between the starlight absorbed by dust and the far infrared emission ([22], [23], [24], [25]) suggest that either the central face-on optical depth of luminous spiral galaxies is more than unity or that the dust emissivity in the sub-millimeter is underestimated by the standard dust model of Draine & Li ([26]) (e.g. [27], [28], [29]). To solve this riddle different methods to estimate the opacity must be employed and the kinematics method might be one of them. The comparison of my result with that of Valotto & Giovanelli ([8]) shows that the uncertainty in the distribution of atomic Hydrogen in spiral galaxies can yield a significant difference of the opacity of spiral galaxies as this is derived from the observed rotation curves. Thus kinematics of the stellar population (e.g. [13]) which is smoothly distributed in the galactic disk are more appropriate for such tasks.